\documentstyle[prl,aps,epsfig,multicol]{revtex}
\newcommand{\beq}{\begin{equation}}
\newcommand{\eeq}{\end{equation}}
\newcommand{\ket}[1]{\left| {#1} \right>}

\newcommand{\cnot}{$^C\!X$~}

\begin{document}

\title{Speed of ion trap quantum information processors}
\author{A. Steane*, C. F. Roos, D. Stevens*, A. Mundt, D. Leibfried,
F. Schmidt-Kaler, R. Blatt}

%EndAName
\address{*Oxford Centre for Quantum Computation, Department of
Physics, University of Oxford, \\ Clarendon Laboratory, Parks
Road, Oxford OX1 3PU, England. \\ Institut f\"ur
Experimentalphysik, Universit\"at Innsbruck, Austria}

\date{\today}
\maketitle

\begin{abstract}
We investigate theoretically the speed limit of quantum gate
operations for ion trap quantum information
processors. The proposed methods use laser pulses for quantum
gates which entangle the electronic and vibrational degrees of
freedom of the trapped ions. Two of these methods are studied in
detail and for both of them the speed is limited by a combination
of the recoil frequency of the relevant electronic transition,
and the vibrational frequency in the trap. We
have experimentally studied the gate
operations below and above this speed limit. In the latter case,
the fidelity is reduced, in agreement with our theoretical
findings.
\end{abstract}

\begin{multicols}{2}

\section{Introduction}

Experimental methods which allow both coherent control, and rapid
and reliable measurement of the quantum state, have been available
for some time for single ions held in ion traps. In an influential
paper, Cirac and Zoller \cite{95:Cirac} showed furthermore how
laser manipulation of a string of ions in a linear ion trap can
achieve general coherent evolution of the joint state of the ions,
using currently available technologies, and with good scaling
properties. Specifically, the coherent evolution can be driven so
as to realize any unitary transformation on the joint state of the
ions, including transitions from product states to entangled
states. Soon after their theoretical work, the essential
ingredients of the method were realized experimentally for a
single trapped ion \cite{95:Monroe}, and more recently two ions
were driven from a product state to an entangled state with around
90\% fidelity using closely related ideas \cite{98:Turchette}.
This is the only efficient creation of entanglement yet observed
in any area of physics.

The combination of universal driven unitary evolution, an exponential
scaling of the available Hilbert space with system size, and reliable
measurement of the resulting state, are the essential ingredients for a
future quantum computer. The scaling of system size in this
argument is measured by the way the whole physical apparatus
becomes larger, and operation of quantum logic gates slower, as
more qubits are added to the system. In the case of the linear ion
trap, the adding of further qubits (ions) into the trap is
straightforward. The growth of the physical apparatus, and the
slowing down of the processing, is dominated by the optical and
electronic equipment needed to control the ions. This growth and
slow-down is a polynomial, not exponential, function of the number
of qubits, for small numbers of qubits. Technical problems will
place a limit, as yet unknown, on the highest number of qubits
which might be feasible, but below this limit the ion trap has the
good scaling properties which allow the essential principles of
quantum computing to be experimentally realised. Hence,
laser-cooled ion trap experiments have a dual interest both for
studying new avenues in atomic physics (e.g. interferometry with
entangled particles) and for understanding quantum information
processing in a specific realisable system.

In this work we examine, both theoretically and experimentally,
the speed with which general processing operations can be driven
in the ion trap system. We are concerned with the intrinsic
limitations imposed by the physics of the system, such as the
tightness of the trap and the presence of a rich energy level
structure in the vibrational modes. The initial proposal of Cirac
and Zoller showed how to drive quantum operations in the limit of
small Rabi frequency (slow operations), and initial experiments
have been in this limit. Our aim here is to clarify the trade-off
between precision and speed of the quantum logic gates, and to
derive the optimal way to operate the processor. This optimum is a
compromise between speed-associated problems such as off-resonant
excitation of unwanted transitions, and the basic decoherence rate
due to environmental coupling which will limit the performance
when the gates are too slow. Our main observation is to confirm
the statement in \cite{97:SteaneB}, that the maximum gate rate
obtained by the Cirac Zoller method scales as the geometric mean
of the trap vibrational frequency and the recoil frequency
associated with the multi-ion string. This is in contrast to
statements made elsewhere that the maximum gate rate is roughly
proportional to the vibrational frequency, though this faster rate
may be available if further tricks are adopted, such as excitation
in the node of a laser standing wave \cite{95:Cirac}. We extend the
discussion in \cite{97:SteaneB} by making a quantitative
statement, equation (\ref{TSmin}), of the relation between speed
and precision of the gate.

We also calculate the maximum gate rate for the method proposed by
Monroe {\em et al.} \cite{97:Monroe}, where a controlled-not
operation is achieved without the need for an additional
transition and laser frequency, by driving multiple Rabi cycles on
the carrier, for well-chosen values of the Lamb-Dicke parameter.

We experimentally study the Cirac-Zoller method, using a single
trapped Calcium ion cooled to the ground state of the motion in
one dimension. The experiments are akin to those in
\cite{95:Monroe,99:Roos} where the methods were first demonstrated
for one ion, except that we examine the regime as yet unexplored,
where the gate rate is above the recoil frequency, so that the
calculations carried out in the small Rabi frequency limit are no
longer appropriate.

The paper is set out as follows. Section \ref{s_theory} presents
the theory of the logic gates, and calculations of their fidelity
as a function of the main parameters, primarily the Rabi frequency
of the atom--light interaction, and the trap tightness. The
calculations involve numerical solution of the Schr\"odinger
equation for the system. We note the need to tune the laser light
to the resonant frequency correctly, that is, taking into account
the light shift (a.c. Stark shift) of the levels. We present
results in the first instance considering only one mode of
vibration of the ion string, and then in section \ref{s_modes}
we briefly discuss the influence of the other modes. Section
\ref{s_exp} presents our experiments. We cool a single trapped
calcium ion to the ground state of the motion in one dimension,
and then observe Rabi flopping on the first blue motional sideband
of the narrow 729 nm transition. By driving the sideband in the
regime where the Rabi frequency is of the order of the trap
vibrational frequency, we observe the expected trade-off between
speed and precision of the operations, caused by off-resonant
excitation of unwanted motional states.

\section{Theory of switching rate} \label{s_theory}
\subsection{Preliminaries}  \label{s_prelim}

Consider a string of two or more ions in a linear trap. The trap
is strongly confining along $x$ and $y$ directions, and less
strongly confining along the $z$ axis. We will assume motion in
$x$ and $y$ directions is unexcited, and consider one normal mode
of vibration of the ion string along $z$, writing its angular
frequency $\omega _{z}$. The other normal modes along $z$ will be
assumed to be unexcited throughout. (We will consider the element
of approximation introduced by this assumption at the end.) For
brevity, we will refer to this vibrational degree of freedom as
``the normal mode'' or ``the vibration''. It has an evenly spaced
ladder of energy levels $E_{n}=(n+1/2)\hbar \omega _{z}$.
Typically in an experiment one would choose the mode of interest
to be the second or third, i.e. $\omega _{z}=\sqrt{3}\,\omega _{z,
\rm cm}$ or $\sqrt{29/5}\,\omega _{z, \rm cm}$, where $\omega _{z,
\rm cm}$ is the frequency of the centre of mass mode. For modes
other than the centre of mass, the Lamb-Dicke parameter describing
the ion--light coupling will vary from one ion to another
\cite{98:James}, but this presents no problem as long as it is
taken into account when choosing laser pulse intensities and/or
durations. We will assume that the laser light is directed onto
one ion at a time.

The Lamb-Dicke parameter is $\eta =\eta _{1}/\sqrt{N}$ where $N$ is the
number of ions, and $\eta _{1}=\sqrt{E_{R}/\hbar \omega _{z}}$. $%
E_{R}=(r\hbar k_{z})^{2}/(2M)$ is the recoil energy for a single
ion initially at rest undergoing a $\pi$ pulse interaction with
the laser field; $k_{z}$ is the $z$ component of the wavevector
${\bf k}$ of the laser field, $M$ is the mass of the ion, and
$r=1$ for a single-photon transition, $r=2$ for a Raman transition
assuming the geometry $k_{1z}=-k_{2z}$ for the two Raman beams.
Example recoil frequencies $E_{R}/h$ are given in table
(\ref{t_tran}).

We will assume throughout that $\eta < 1$. This is the regime in
which the ion trap processor is used in practice, both because it
facilitates the initial cooling to the ground state of motion, and
because the processor then runs faster, as we will discuss.

\end{multicols}

\begin{table}
\begin{tabular}{llllr@{}l}
ion & mass & qubit basis & transition & recoil frequency (kHz)\\
    &      &             &            &  \\
Beryllium & 9  & hyperfine ground states F=2 F=1 & 313\,nm Raman &
452& \\ Calcium  & 40 & Zeeman ground states S$_{\pm 1/2}$ &
397\,nm Raman & 63& \\ Calcium  & 40 & S$_{1/2}$ ground and
D$_{5/2}$ metastable state & 729\,nm single-photon & 4.7 &
\end{tabular}
\caption{Relevant parameters for the implementation of quantum
information in trapped ions: wavelength, recoil frequencies for
qubit candidate ions, assuming $45^{\circ}$ angle between laser
wave vector and the $z$ axis.} \label{t_tran}
\end{table}

\begin{multicols}{2}

The Cirac-Zoller method adopts the internal state of each ion as a
qubit (we restrict attention to two internal states and thus one
qubit per ion). Gates between qubits are obtained via excitation
of the common vibrational mode, and measurement of the state of
one or more qubits is by observing fluorescence. The method uses
the fact that arbitrary single-qubit rotations, combined with any
one 2-qubit gate such as controlled-not ($^C\!X$) or
controlled-rotation ($^C\!Z$), between arbitrary pairs of qubits,
form a universal set \cite{85:Deutsch,96:Barenco,98:Steane}. That
is, any unitary transformation can be decomposed into elements
from the set. In the trapped ion system, a single-qubit rotation
is a transition in the internal state of a single ion, driven by a
laser pulse. A $^C\!X$ between the internal states of any pair of
ions $A$ and $B$ is achieved by a swap ($S$) operation between the
internal state of $A$ and the vibration, followed by $^C\!X$
between the vibration and the internal state of ion $B$ (with
vibration as control, ion internal state as target) followed by
$S$ again on ion $A$. Therefore the only operations that will
concern us are single-qubit rotations, and $S$ and $^C\!X$ between
a single ion and the vibration. All of these are achieved by
excitation of a chosen transition. The unwanted off-resonant
excitation of other transitions are the main subject of this
paper.

The strength of the ion-laser interaction is parametrized by the
Rabi frequency, given by

\begin{eqnarray}
\Omega_{nm} &=& \left< {n} \right| \exp \left( i \eta
(\hat{a}^{\dagger} + \hat{a}) \right) \left| {m} \right>
\Omega_{{\rm free}} \\ &\equiv& C_{nm} \Omega        \label{Omega}
\end{eqnarray}

where $\Omega_{{\rm free}}$ is the Rabi frequency for a free ion,
the states $\left| {n} \right>$ are vibrational energy
eigenstates, $\Omega = \exp(-\eta^2/2) \Omega_{{\rm free}}$ and
the factor $C_{nm}$ is given by the following equation (3). Values
of $C_{nm}$ are listed in table (\ref{t_C}) for the low-lying
vibrational levels.

\end{multicols}

  \beq
C_{nm} = \sqrt{m! n!} (i\eta)^{|f-m|} \sum_{j=0}^{{\rm min}(m,n)}
\frac{ (-1)^j \eta^{2 j} } {j! (j+|n-m|)! ( {\rm min}(m,n) - j )!
} .  \label{Cnm}
  \eeq

\begin{table}[tbp]
\begin{tabular}{c|cccc}
$C_{nm}$ & 0 & 1 & 2 & 3 \\ \hline 0 & 1                       &
$i \eta$   & $-\eta^2/\sqrt{2}$
  & $-i \eta^3 / \sqrt{6}$ \\
1 & $i \eta$ & $(1-\eta^2)$ & $i \sqrt{2} \eta (1 - \eta^2/2)$
  & $-\sqrt{3/2}\,\eta^2 (1- \eta^2/3)$ \\
2 & $-\eta^2/\sqrt{2}$      & $i \sqrt{2} \eta (1 - \eta^2/2)$
  & $(1 - 2 \eta^2 + \eta^4/2)$
  & $i \sqrt{3} \eta (1 - \eta^2 + \eta^4/6)$ \\
3 & $-i \eta^3 / \sqrt{6}$  & $-\sqrt{3/2}\,\eta^2 (1- \eta^2/3)$
  & $i \sqrt{3} \eta (1 - \eta^2 + \eta^4/6)$
  & $1 - 3\eta^2 + 3 \eta^4/2 - \eta^6/6$
\end{tabular}
\caption{Matrix element for vibrational-state-changing
transitions.} \label{t_C}
\end{table}

\begin{multicols}{2}

The single-qubit rotations can be much faster than the two-qubit
gates, because the energy separation $\hbar \omega_0$ of the
internal energy levels is much larger than that of the vibrational
levels, and by choosing a laser beam direction $k_z = 0$ (single
photon transition) or $k_{1z} = k_{2z}$ (Raman transition), the
Lamb-Dicke parameter can be made to vanish during single-qubit
operations, which means these operations do not couple to the
vibrational state ($C_{nm}$ becomes $\delta_{nm}$). Therefore, for
$\Delta n = 0$, $\Omega$ can be large compared to $\omega_z$
without causing off-resonant excitation of $\Delta n \ne 0$
transitions. Therefore, the speed of the ion trap processor is
limited by the $S$ and $^C\!X$~ gates.

The $S$ gate is achieved by a $\pi$ pulse on either the first red
vibrational sideband, that is, at frequency $\omega_0 - \omega_z$,
or the first blue vibrational sideband at frequency $\omega_0 +
\omega_z$. The duration of the $S $ gate is therefore
\begin{equation}
T_S = \frac{\pi}{\eta \Omega}.  \label{Tswap}
\end{equation}
where the gate becomes exact in the limit $\Omega \rightarrow 0$.
The choice of red or blue sideband can be dictated by experimental
convenience. The two are equivalent in terms of their quantum
computational effect, since the logical operations produced can be
made identical simply by relabelling the states (i.e. changing
which internal state of the ion is called 0, and which is called
1). We will treat the red sideband throughout our theoretical
discussion, but the results will apply equally to blue sideband
excitation.  In fact, we used a blue sideband in the experiments
described in section \ref{s_exp}.

We will consider two methods for the \cnot gate. The first is that
described in the original proposal of Cirac and Zoller, where
$^C\!X$ is obtained by single-bit rotations (Hadamard gates)
combined with $^C\!Z$, and $^C\!Z$ is obtained from a $2\pi$ pulse
on the first red sideband (or blue, depending on the relative
positions of the levels) of an auxiliary transition in the ion,
i.e. laser frequency $\omega_{\rm aux} - \omega_z$, gate duration
\beq T_{C1} = \frac{2 \pi}{\eta \Omega}.  \label{Tcnot1} \eeq

The other method is that of Monroe {\em et al.} \cite{97:Monroe},
who proposed using a $2m \pi$ pulse on the carrier (frequency
$\omega_0$) where $m$ is an integer and $\eta^2 = 1/(2 m)$. This
is especially useful if an auxiliary transition is not available.
When the vibrational state is $\ket{n=0}$, this drives an integer
number of Rabi oscillations of the internal state, while if the
vibrational state is $\ket{n=1}$, this drives a half-integer
number of Rabi oscillations, since $C_{11} = 1 - \eta^2 =
(2m-1)/(2m)$. The method requires $\eta$ to be fixed in the
experiment to one of the special values $1/\sqrt{2 m}$, which is
easily done in practice by adjusting the trap confinement and/or
laser beam direction.  The duration of a \cnot gate by the Monroe
method is \beq T_{C2} = \frac{2 m \pi}{\Omega} = \frac{\pi}{\eta^2
\Omega}.  \label{Tcnot} \eeq Since the ion trap is operated with
$\eta < 1$, it might be thought that $T_{\rm C2}$ is necessarily
greater than $T_{\rm C1}$. In fact, this is not necessarily the
case, since these gate times are limited by the maximum allowable
Rabi frequency $\Omega$, and this can depend on the type of gate.

Sorensen and Molmer \cite{99:Sorensen,99:Molmer} have proposed
a general method to implement gates such as controlled-not using bichromatic
laser fields, achieving good fidelity even when the vibrational degrees of
freedom are not in their ground state. The speed limitations of this
method have recently been re-examined \cite{00:Sorensen}, so we will not
discuss it here, other than to say the method appears to be useful.

\subsection{Solution of time-dependent Schr\"odinger equation}

Our aim is to find the maximum switching speed of the processor. It is seen
from equations (\ref{Tswap}), (\ref{Tcnot1}) and (\ref{Tcnot}) that this is
determined by the maximum allowable Rabi frequency during the gates which
operate on the vibrational state. The Rabi frequency cannot be arbitrarily
large, because in the limit $\Omega \gg \omega_z$, the evolution would be
independent of the vibrational state.

In order to find the maximum Rabi frequency, we need to solve the
time-dependent Schr\"odinger equation for the system, without making the
approximation of small $\Omega$. We assume a two-level ion. This means we
will not explicitly examine the type of $^C\!X$ gate which uses an auxiliary
level (equation (\ref{Tcnot1})), but in any case this operation is closely
related to the $S$ operation which we will examine, so results for the rate
of $S$ will apply to this type of $^C\!X$ apart from the factor 2. We will
study $S$ and the Monroe $^C\!X$ gate.

The Hamiltonian for the ion that is being illuminated by the laser during a
given gate is
\begin{equation}
H = H_0 + H_I
\end{equation}
where $H_0$ is diagonal, with diagonal elements given by $\hbar ( 0,
\omega_z, 2 \omega_z, \cdots, \omega_0, \omega_0 + \omega_z, \omega_0 +
2\omega_z, \cdots )$, and the interaction Hamiltonian
\begin{equation}
H_I = \hbar \frac{\Omega}{2} \left(
\begin{array}{cc}
{\bf 0} & \hat{C} e^{i \omega_L t} \\
\hat{C}^{\dagger} e^{-i \omega_L t} & {\bf 0}
\end{array}
\right)
\end{equation}
where $\hat{C}$ is the matrix having elements $C_{nm}$. Note that the only
approximation so far is to ignore the other vibrational modes.

We first adopt a frame rotating with the laser frequency:
$
\left| {}\right. \!{\tilde{\psi}(t)}\left. \!\right\rangle =U\left| {\ \psi
(t)}\right\rangle,
$
where $U$ is a diagonal unitary matrix, with diagonal elements $\exp
(i\{-1, -1, -1, \cdots , 1, 1, 1, \cdots \}\omega _{L}t/2)$. The states $\left|
{}\right. \!{\tilde{\psi}(t)}\left. \!\right\rangle $ satisfy the Schr\"{o}%
dinger equation $i\hbar (d/dt) \left| {\tilde{\psi}} \right\rangle
= \tilde{H}\left| {\tilde{\psi}} \right\rangle $, where
$
\tilde{H}\equiv i\hbar (dU/dt) U^{\dagger }+UHU^{\dagger }
\label{Htilde}
$
is now a time-independent Hamiltonian. We can therefore write the solution
to the Schr\"{o}dinger equation
\begin{equation}
\left| {}\right. \!{\tilde{\psi}(t)}\left. \!\right\rangle =e^{-i\tilde{H}%
t/\hbar }\left| {}\right. \!{\tilde{\psi}(0)}\left. \!\right\rangle \equiv
V^{\dagger }e^{-iV\tilde{H}V^{\dagger }t/\hbar }V\left| {}\right. \!{\
\tilde{\psi}(0)}\left. \!\right\rangle
\end{equation}
where $V$ is the matrix of eigenvectors of $\tilde{H}$, so that $V\tilde{H}%
V^{\dagger }$ is diagonal.

The quantity of interest, from the point of view of quantum information
processing in the trap, is the final state expressed as a superposition of
computational basis states. If at $t=0$ the computational basis states are $%
\left| {\tilde{u}(0)} \right> = \left| {u(0)} \right>$, then at other times
they are $\exp(-i \tilde{H}_0 t/\hbar) \left| {\tilde{u}(0)} \right>$ since
then the coefficients $\left< {\tilde{u}(t)} \right. \left| \right. \! {%
\tilde{\psi}(t)} \left. \! \right>$ do not evolve in the absence of gate
operations. Therefore we would like to calculate
\begin{equation}
\left< {\ u(0) } \right| e^{-i \left(\tilde{H} - \tilde{H}_0 \right) t /
\hbar} \left| {\ \psi(0) } \right>.  \label{calc}
\end{equation}

Define $P \equiv \exp(-i(\tilde{H} - \tilde{H}_0)t/\hbar)$, and let $G$ be
the precise unitary operator for the intended gate. Then the degree to which
the laser pulse produces the intended gate is given by the overlap between
the final state and one which would be obtained from $G$. To indicate the
degree of imperfection of the laser pulse operation, we therefore calculate
\begin{equation}
f_{{\rm min}} = \min_{\psi} \left|\left< {\psi} \right| G^{\dagger} P \left|
{\psi} \right>\right|^2  \label{fmin}
\end{equation}
and define the imprecision to be $\epsilon = (1 - f_{{\rm min}})^{1/2}$.

Let us comment on whether or not $\epsilon \ne 0$ represents imperfection in
the system. We are assuming no imperfection in the sense of unknown dynamics
(e.g. laser intensity and frequency noise). Therefore as long as we can
calculate numerically the effect of the laser pulse operation, we have
accurate knowledge of the expected state of the system: the fact that the
laser pulse gate differs from any particular ``ideal'' gate is not a source
of any imprecision at all. However, the intention is to use the ion trap as
a quantum processor, to do quantum calculations which we lack the computing
power to simulate classically. Putting together many laser pulses, we are
therefore assuming we cannot predict the effect on the whole quantum process
of having $\epsilon \ne 0$ in all the operations. Therefore $\epsilon$ must
be regarded as imprecision in the device, which must be minimized.

The question of driving complicated evolution in a predictable manner is
central to other techniques such as NMR spectroscopy, and there exist
methods to combine pulses causing different types of driven rotation in order to
undo the effect of certain terms in the Hamiltonian. Equivalent methods can
almost certainly be found for the ion trap system, but in any case
operations close to ideal ones will remain the best starting point for any
such method.

\subsection{Performance without correction for light shift}

In this section, we will examine the operation of the processor
in the regime $\eta < 1$, in the case that the light shifts (a.c.
Stark shifts) are not taken into account when choosing the laser pulse
frequency, phase, and duration.

First consider the $S$ gate: a $\pi$ pulse on the first red sideband. We set
$\omega_L = \omega_0 - \omega_z$ in (\ref{Htilde}), and examine the
eigenvalues of $\tilde{H}$. The separations between the eigenvalues enable
one to deduce the Rabi flopping frequencies and the light shifts in the
system. The Rabi flopping frequency on the $\left| {n=0} \right>
\leftrightarrow \left| {n=1} \right>$ transition is $\eta \Omega$, as
expected, which confirms that the gate time is $T_{S} = \pi/(\eta \Omega)$ as
in equation (\ref{Tswap}).

The frequency of the transition is shifted by
\begin{equation}
\Delta \omega = \frac{1}{2} \left( 1 + \frac{\eta^2}{2} \right) \left(\frac{
\Omega }{\omega_z} \right)^2 \omega_z + \frac{1}{8} \left(\frac{ \Omega }{%
\omega_z} \right)^4 \omega_z + \cdots.  \label{dos}
\end{equation}
This is primarily the light shift caused by the presence of $\Delta n = 0$
transitions, which are off-resonant by $\omega_z$.

As was noted by Wineland {\em et al.} \cite{98:WinelandB}, the
primary error in the gate operation is caused by the light shift
$\Delta \omega$, which can be significant compared to the Rabi
flopping frequency $\eta \Omega$. To understand its effect, we now
model the transition of interest, that is $\left| {g,n=1} \right>
\leftrightarrow \left| {e,n=0}
\right>$, as a two-level system, driven by a $\pi$ pulse off-resonant by $%
\Delta \omega$. The only term in the propagator of first order in $\Delta
\omega$ is $i (\Delta \omega/\alpha) \sin(\alpha t/2)$ where $\alpha^2 =
\eta^2 \Omega^2 + \Delta \omega^2$. Putting $\eta \Omega t = \pi$, we obtain
\begin{equation}
\epsilon \simeq \left| \frac{i \Delta \omega}{\alpha} \right| \simeq \frac{%
\Omega}{2 \eta \omega_z}
\end{equation}
Therefore, to attain a given gate precision $\epsilon$ the Rabi frequency
must satisfy $\Omega \le 2 \epsilon \eta \omega_z$, which gives
\begin{eqnarray}
\frac{1}{T_S} = \frac{\eta \Omega}{\pi} &\le& \frac{2 \epsilon}{\pi} \eta^2
\omega_z \\
&=& 4 \epsilon \frac{E_R}{N h}.  \label{S1}
\end{eqnarray}
The result thus has the simple form that the processor speed for swap
operations is limited by the $N$-ion recoil frequency.

Next, consider the $^C\!X$gate using the special Lamb-Dicke parameter method
of Monroe {\em et al.}, that is, a $2 m \pi$ pulse on the carrier with $2 m =
1 / \eta^2$. Substituting $\omega_L = \omega_0$ into (\ref{Htilde}), and
examining the eigenvalues of $\tilde{H}$, we find the transition frequencies
are shifted by
\begin{equation}
\Delta \omega \simeq \frac{ (\eta \Omega)^2 }{\omega_z}. \\
\end{equation}
Since $\Delta \omega \ll \Omega$, the light shifts for this gate are much
less significant. Modelling each driven transition as a two-level system,
using the same method described above for the $S$ gate, we find that the
effect of $\Delta \omega$ is $\epsilon \simeq \eta^2 \Omega/\omega_z$. This
is small compared to the effect of off-resonant excitation of $\Delta n =
\pm 1$ transitions. These are driven in the regime where the detuning is
large compared to their Rabi frequency, leading to a propagator with terms
of order $\sqrt{2} \eta \Omega/ \omega_z$, so we find
\begin{equation}
\epsilon \simeq \frac{\sqrt{2} \eta \Omega}{\omega_z}.  \label{epsC}
\end{equation}
Setting $\Omega \le \epsilon \omega_z / \sqrt{2} \eta$, we find
\begin{equation}
\frac{1}{T_{C2}} \le \sqrt{2} \epsilon \sqrt{\frac{E_R}{N h} \frac{\omega_z}{%
2 \pi} }, \;\;\;\;\;\;\;\Omega \ll \omega_z.  \label{TC2min}
\end{equation}
It might appear from this that the gate speed can increase without limit by
using a tight trap, but this is a false impression because the expression is
only valid in the regime $\Omega \ll \omega_z$, which means it is only valid
for a gate rate small compared to $\eta^2 \omega_z / \pi \simeq E_R/(N h)$.

\subsection{Gates with correction for light shift: Red sideband} \label{redcorrect}

The light shift is readily corrected for by tuning the laser
frequency to accurate resonance with the light-shifted transition
frequency \cite{98:WinelandB}. We will now describe the correct
laser pulse frequency, phase, and duration, and derive an
expression for the imprecision caused by the primary remaining
unwanted effect, which is off-resonant excitation.  We find that
the recoil frequency still strongly influences the processor
speed.

We have studied this problem by numerically evaluating expressions (\ref
{calc}) and (\ref{fmin}). Only vibrational levels with $n < 4$ were included,
thus yielding an $8 \times 8$ propagator matrix. When further vibrational
levels were included in the calculation, the results were not significantly
affected.

We find that the best implementation of a swap gate is obtained when the
laser frequency is tuned to the light-shifted frequency of the first red
sideband: $\omega_L = \omega_0 - \omega_z + \Delta \omega$ where $\Delta
\omega$ is given by equation (\ref{dos}). The gate time is still as in
equation (\ref{Tswap}). With no further adjustments, this produces a
propagator close to $U_{\phi} S$ where $S$ is a perfect swap operator,
and $U_{\phi}$ is diagonal with diagonal
elements $\exp( i \{ 1, 1, 1, \cdots, -1,
-1, -1, \cdots \} \Delta \omega T_S /2 )$. Therefore, to produce the
intended gate, the system must be further evolved by an application of $%
U_{\phi}^{\dagger}$. This is equivalent to applying subsequent laser pulses
with a correspondingly adjusted phase. Therefore the swap gate is completed
by changing the phase of the oscillator used in the experimental apparatus
to keep the laser in step with the atom's internal resonance (at
frequency $\omega_0$), by an amount $\Delta \phi = \Delta \omega T_S$.

This adjustment permits the swap gate rate to exceed the recoil frequency.
The term in the system propagator $P$ which now gives the dominant
contribution to the imprecision $\epsilon$ is off-resonant excitation
of $\Delta n = 0$ transitions. We expect the
amplitude for this process to be of
order $\Omega / \sqrt{ \Omega^2 + \omega_z^2 }$. Our numerical calculations
confirm this, giving
\begin{equation}
  \epsilon \simeq \frac{\Omega}{\sqrt{2 (\Omega^2 + \omega_z^2)}} \simeq
  \frac{\Omega}{\sqrt{2}\, \omega_z}   \label{epsS}
\end{equation}
for the remaining imprecision in the adjusted swap operation. Therefore the
gate rate is now limited by
\begin{equation}
\frac{1}{T_S} \le 2 \sqrt{2} \, \epsilon \sqrt{\frac{E_R}{N h}
\frac{\omega_z}{2 \pi} },                  \label{TSmin}
\end{equation}
which is $2 \sqrt{2} \, \epsilon$ times the geometric mean of the $N$-ion
recoil frequency and the vibrational frequency of the chosen normal mode.

If an auxilliary transition is available, the above argument will hold for
the $^C\! Z$ gate implemented by a $2 \pi$ pulse to the first vibrational
sideband of the auxilliary transition, the only difference is that the gate
rate at given $\epsilon$ is half that for the {\sc swap} gate ($T_{C1} = 2
T_{S}$).

\subsection{Carrier}

The Monroe method $^C\!X$~ gate is executed by a laser pulse at the carrier
frequency. The speed limit given by equation (\ref{TC2min}) cannot be exceeded,
since it is caused by off-resonant excitation that cannot be avoided.
However, we are interested in the behaviour outside the region of validity
of (\ref{TC2min}), that is, when $\Omega \sim \omega_z$.

When $\eta$ is small but $\Omega$ is not, the precision of the operation need
not be limited by off-resonant excitation of $\Delta n = \pm 1$ transitions,
since their rate is low. The source of imprecision instead comes from the
fact that the dynamics of the system can no longer be pictured as separate
Rabi flopping on the $n=0 \leftrightarrow 0$ transition and the $n=1
\leftrightarrow 1$ transition. Rather, the interaction Hamiltonian drives the
set of vibrational levels $n=0,1,2, \cdots \sim \Omega/\omega_z$ as a single
entity. The result is that the evolution from an initial state $\left| {n=0}
\right>$ is almost the same as the evolution from an initial state $\left| {%
n=1} \right>$. Therefore the Rabi flopping must be driven for a longer time
before a $^C\!X$~ gate is obtained, and the gate speed ceases to increase
with $\Omega$. The rate obtained at $\Omega \sim \omega_z$ was found to be
the maximum. Recalling the expression (\ref{Tcnot}) for $T_{C2}$, it is seen
that this maximum is approximately the recoil frequency.

In the region $\Omega \sim \omega_z$, the gate can be optimized simply by
adjusting the pulse duration. The correct pulse length is not precisely $\pi
/ (\eta^2 \Omega)$, but differs from this by an amount of order $\pi / (\eta
\Omega)$. We have not found any simple expression for this adjustment, we
surmise that this is because it is a complicated function of all the light
shifts in the multi-level system. To run a processor in practice, the
correction to the gate time can either be calculated numerically, or
measured experimentally. By this adjustment, we were able to extend the
region of validity of expression (\ref{TC2min}) up to $\Omega \simeq 0.5
\omega_z$. Operating at this limit, the gate rate is always equal to the $N$%
-ion recoil frequency, independent of the trap tightness, but the gate
becomes more precise as the trap gets tighter ($\epsilon \simeq \eta/\sqrt{2}
$, expression (\ref{epsC})).

One way to use the carrier excitation method to achieve faster $^C\!X$ gates,
is to use higher vibrational levels such as $\left| {n=2} \right>$ (in
conjunction with $\left| {n=0} \right>$). However, this would mean that
the $S$ gate
would have to drive the second red sideband, making it slower (the expected
limitation at small $\eta$ being off-resonant excitation of $\Delta n = 0$
transitions). Therefore only small gains, if any, are available by this route.

\subsection{Sensitivity to laser intensity fluctuations}

Although our main purpose is to study limitations imposed by the
unavoidable properties of the system, we comment here on the sensitivity
to one source of technical noise, namely laser intensity fluctuations,
since these partially limit our experiments described in section \ref{s_exp}.
The laser drives Rabi flopping at frequency $\eta \Omega$ for the
Cirac-Zoller swap gate, and at frequency $\Omega$ for the Monroe gate.
If, owing to intensity fluctuations, $\Omega$ is imprecise by $\Delta \Omega$
then the action of either gate will be imprecise by
$\epsilon \simeq T_S \eta \Delta \Omega / \pi$
and $\epsilon \simeq T_{C2} \Delta \Omega / \pi$ respectively.
We therefore find
$\Delta \Omega / \Omega \simeq \epsilon$ for the swap gate (also
for the closely related $^C\!Z$ gate), whereas
$\Delta \Omega / \Omega \simeq \eta^2 \epsilon$ for the Monroe gate.

\subsection{Allowance for all the normal modes} \label{s_modes}

The only approximation we made was to neglect all but one of the
normal modes of oscillation of the ion string. When we relax this
assumption, the main features of the more general problem can be
understood in terms of the Rabi frequencies and the normal mode
frequencies. To keep the notation clear, we will describe a case
of three normal modes, which we will label $x$, $y$ and $z$, but
the discussion readily generalises to more normal modes. In our
experiments described in section \ref{s_exp} there is a just a
single ion, so the modes of interest are indeed vibration along
the principle axes of the trap, but the modes here labelled $x$,
$y$ and $z$ could equally refer to different normal modes of the
$z$-oscillation of a three-ion string, in the case where the two
other directions of oscillation are frozen out owing to the linear
geometry (that is, they have much higher frequency, and are
prepared in the ground state). Equation (\ref{Omega}) is replaced
by

\end{multicols}

\beq \Omega(n_x,m_x)(n_y,m_y)(n_z,m_z) = C_{n_x m_x}(\eta_x)
C_{n_y m_y}(\eta_y) C_{n_z m_z}(\eta_z) \Omega \eeq

\begin{multicols}{2}

in an obvious notation, where we now redefine $\Omega \equiv
\exp(-(\eta_x^2 + \eta_y^2 + \eta_z^2)/2) \Omega_{\rm free}$.
There are two regimes to consider. In the case that all the modes
are prepared very close to the ground state, the main effect is
that the light shift is no longer that given by equation
(\ref{dos}), but one given by a sum of light shifts related to all
the possible transitions. This is easily compensated by a suitable
adjustment to the laser frequency. The limit on precision
therefore remains that due to off-resonant transitions. For
excitation on the red sideband ($n_z = 0, m_z=1$), in addition to
carrier transitions excited off-resonant by $\omega_z$ with Rabi
frequency $\Omega$, there are also sideband transitions excited
off-resonant by $\omega_x - \omega_z$ and $\omega_y - \omega_z$,
with Rabi frequencies $\eta_x \Omega$ and $\eta_y \Omega$
respectively. Since the latter contribute at a higher order in
$\eta$ compared to the carrier term, they are negligible unless
two modes are close in frequency. For a linear ion string the
separation of the lowest mode frequencies is large enough that the
additional terms make little contribution. Therefore our previous
discussion, including equations (\ref{epsS}) and (\ref{TSmin}),
remains valid. In the experiments described in section
\ref{s_exp}, on the other hand, $\omega_y - \omega_z \simeq
0.025 \,\omega_z$ so unwanted excitation of the $y$ sideband has
to be taken into account.

For the carrier excitation, in addition to the
$z$ sidebands excited off-resonant by $\omega_z$ with Rabi frequency
$\eta_z \Omega$, there are also sideband transitions excited off-resonant
by $\omega_x$ and $\omega_y$ with Rabi frequencies
$(1-\eta_z^2) \eta_x \Omega$ and $(1-\eta_z^2) \eta_y \Omega$ respectively.
These contribute at the same order in $\eta$ as the $z$ sidebands, and
result in a change to equation (\ref{TC2min}) by a numerical factor of order unity.

Our experiments were carried out with only one mode cooled to the
ground state. The other two modes had roughly thermal
distributions with mean vibrational quantum number of order 10.
The result of this is to blur all the Rabi frequencies and light
shifts, since each time a given $z$ transition is excited, the
Rabi frequency depends on the $x$ and $y$ vibrational quantum
numbers:

\end{multicols}

\beq \Omega(n_x, n_x)(n_y, n_y)(n_z, m_z) \simeq \left(1-n_x
\eta_x^2\right)\left(1-n_y \eta_y^2\right) C_{n_z m_z} \Omega.
\label{Oxy} \eeq

\begin{multicols}{2}

We are able to fit our experimental results by calculating the
evolution for each value of $n_x, n_y$, and then averaging over a
thermal population of the $x$ and $y$ vibrations.

\subsection{Summary}

In conclusion, we find that the swap gate and the Cirac-Zoller
type of controlled-phase gate can be made faster, at a given level
of precision, by making the trap tighter, but the speed increase
is only in proportion to $\omega_z^{1/2}$. This is in agreement
with earlier work \cite{97:SteaneB}, but here we have added a
precise quantitative statement of the trade-off between speed and
precision. Some previous studies have adopted $T_S \sim \pi /
\omega_z$ for the purpose of making rough estimates, but we find
the error thus introduced is significant; for example, it leads to
an overestimate of the switching rate by two orders of magnitude
in \cite{96:Hughes}. By exciting the ion in the node of laser
standing wave, it is possible to avoid (or greatly reduce) the
off-resonant carrier excitation \cite{95:Cirac,98:James},
therefore $T_S \sim \pi / \omega_z$ may be available, but this
involves a significant extra complication of the experiment which
may render it unfeasible in practice \cite{comment}.

The ``magic Lamb-Dicke parameter" method of Monroe {\em et al} can
not be speeded up indefinitely, it has a natural speed limit given
by the recoil frequency.

There are advantages in having the ions spaced by a micron or
more, in order to allow laser addressing of one ion at a time,
and/or resolving the fluorescence from different ions with
appropriate imaging optics \cite{99:Nagerl}. If we set a limit $s$
for the smallest permissible seperation of the closest ions in the
string during processing, we obtain \cite{96:Hughes,97:SteaneB}.
\beq \omega_{z, \rm cm}^2 < \frac{8 e^2}{4 \pi \epsilon_0 M s^3
N^{1.71}} \eeq Putting this in equation (\ref{TSmin}), and
adopting the breathing mode $\omega_z = \sqrt{3} \omega_{z, \rm
cm}$, we obtain for the fastest swap gate rate, at given precision
and ion spacing, \beq \frac{1}{T_S} \simeq 2.5 \left( \frac{e^2}{4
\pi\epsilon_0} \right)^{1/4} \epsilon \left( \frac{E_R}{h}
\right)^{1/2} \frac{1}{M^{1/4} s^{3/4} N^{0.93}}. \label{TSmm}
\eeq  If we approximate the $N$ dependence as $N^{-1}$, and set $s
= 10 \lambda$ where $\lambda$ is the laser wavelength, then we
arrive at a ``gate time per ion" which depends only on the choice
of ion and transition, at given $\epsilon$. This quantity, $T_S /
N$, is given in table~\ref{t_rate} for some example transitions,
for the case of 99\% gate fidelity (i.e. $\epsilon = 0.1$).

\begin{table}
\begin{tabular}{lcccc}
ion & mass & wavelength & $E_R/h$ (kHz)  & gate time per ion \\
Beryllium & 9 & 313 nm & 452 &  1.26 $\mu$s \\ Calcium  & 40 & 397
nm & 63  &  5.6 $\mu$s \\ Calcium  & 40 & 729 nm & 4.7 &  34
$\mu$s
\end{tabular}
\caption{Minimum gate time per ion for 99\% fidelity {\sc swap} gates,
for three example transitions. The gate time per ion, $T_S / N$,
is given by equation (\protect\ref{TSmm}), for the case $\epsilon=0.1$,
$s = 10 \lambda \simeq 3, 4, 7\;\mu$m.}
\label{t_rate}
\end{table}

\section{Experiments}  \label{s_exp}

In this section, we describe our experimental investigations, carried out in
conditions where the Rabi frequency
$\Omega_{\rm sideband} \equiv \eta \Omega_{\rm carrier}$
of the sideband excitation
is significantly above the recoil frequency. We thus are able to investigate
the regime where the speed of execution of the gates
determines the fidelity of operation.

We use a single trapped $^{40}$Ca$^+$ ion to demonstrate the
principle of quantum information processing. The electronic
$|S_{1/2}$, $m = -1/2\rangle$ ground state and the metastable
$|D_{5/2}$, $m = -5/2\rangle$ state (1s lifetime) are used to
implement one qubit. We apply a 4 Gauss bias magnetic field to
lift the degeneracy of sublevels in the ground and excited state
manifolds. The qubit can be coherently manipulated by laser light
at 729~nm (see table \ref{t_tran}). Two basic operations are
demonstrated. First there is the single qubit rotation which only
affects the individual internal electronic state, and secondly we
perform the swap operation which entangles the electronic state of
the ion and its vibrational state. We did not investigate the
Monroe {\em et al.} $^C\!X$ gate because our experiments involved
small Lamb Dicke parameters of order $0.045$. In this regime the
Monroe method is $\sim 500$ times more sensitive than the
Cirac-Zoller method to laser intensity noise and thermal
populations in spectator modes, and we found it to be unworkable
in these experiments.

We store the ion in a conventional spherical-quadrupole Paul trap
with trap frequencies $\omega_{x, y, z}=2 \pi (4.0, 1.925,
1.850)\,$MHz. To acquire each experimental data point, at given
values of the parameters, we run a series of 100 cycles. A single
experimental cycle is made up of 5 stages, which are (i) Doppler
cooling, (ii) sideband cooling, (iii) coherent driven evolution,
(iv) observation of fluorescence, (v) deshelving. We then record
the fraction $P_D$ of the 100 cycles in which fluorescence was
observed in the penultimate stage.

In more detail, the 5 stages of a cycle are as follows. Doppler
cooling is performed on the $|S_{1/2}\rangle$ to $|P_{1/2}\rangle$
397~nm electric dipole transition. The electronic ground state is
then prepared in a pure $|S_{1/2}, m=-1/2\rangle$ state by optical
pumping. The ion is cooled to the vibrational ground state of the
$\omega_z$ mode by applying sideband cooling on the $|S_{1/2},
m=-1/2 \rangle \leftrightarrow |D_{5/2}, m=-5/2 \rangle $ electric
quadrupole transition at 729~nm. A full description of the trap
and the cooling procedure is given in Ref. \cite{99:Roos}. The
wavevector ${\bf k}$ of the 729~nm radiation is nearly
perpendicular to the $y$ direction, and has an angle of
40$^{\circ}$ and 50$^{\circ}$ to the $z$ and $x$ directions. The
corresponding values of $\eta_{x,y,z}^{729nm}$ are (0.04, 0.01,
0.045). In the qubit operation step, the $|S_{1/2}, m=-1/2 \rangle
\leftrightarrow |D_{5/2}, m=-5/2 \rangle $ transition is excited
with a laser pulse of well controlled frequency, intensity, and
timing. Then, by monitoring fluorescence at 397~nm under laser
excitation, we detect whether a transition to the non-fluorescing
state D$_{5/2}$ occurred. The scheme allows one to discriminate
between the internal states of the ion with an efficiency close to
100\% \cite{86:Sauter,86:Nagourney,86:Bergquist}. In our
experiment,the discrimination effiency is approximately 99.8$\%$,
limited by the lifetime of the metastable state \cite{Roos2000}.

Finally, the ion is again repumped (deshelved)
from the D$_{5/2}$ level to the electronic ground state via the P$_{3/2}$ level.
The fraction $P_D$ of cycles in which fluorescence is observed
indicates the population of the $D_{5/2}$ level after the coherent qubit
operation step.

\subsection{Carrier excitation}

As discussed in section \ref{s_prelim}, the single qubit rotation
can be driven fast and without loss in contrast, even in the range
where $\Omega$ exceeds the trap frequencies $\omega_{x,y,z}$, as
long as the Lamb-Dicke parameters are sufficiently small. To reach
this limit in the case of a single-photon transition under
travelling wave excitation, the wavevector ${\bf k}$ of the
exciting light field should be perpendicular to the chosen
vibration direction, and to the direction having next smallest
vibrational frequency. The remaining vibrational frequency must be
large compared to $\Omega$. In our case the laser is nearly
perpendicular to one of the weak confinement directions, but not
the other, so there is some unwanted excitation of the motion. A
limit to execution speed is set by the available laser power to
drive the qubit transition. However, Rabi frequencies as high as a
few MHz can be achieved even in the case of dipole forbidden
quadrupole transitions.

\begin{center}
\begin{figure}[tbp]
\epsfxsize=0.95\hsize \epsfbox{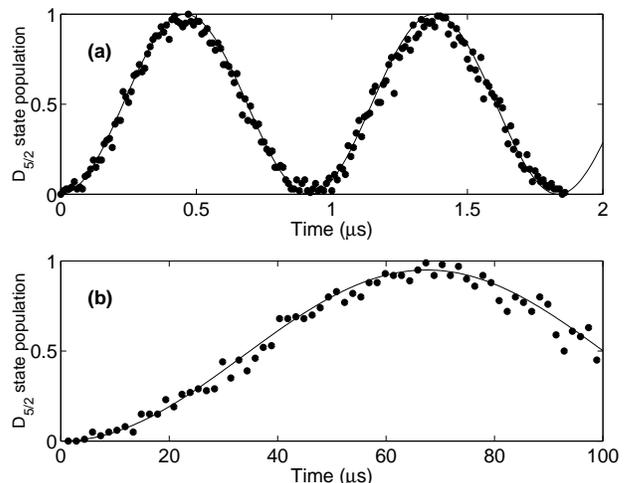} \caption{ a) Rabi
oscillations on the carrier. We excite the $\Delta n=0$ transition
with a laser power of 100~mW and observe a $\Omega =
(2\pi)\,1090$\,kHz Rabi frequency. A contrast of better than 95\%
is reached for $\pi$ and $2\pi$ pulses. b) Rabi oscillations on
the blue sideband at 4mW excitation power. At this low power the
contrast for $\pi$ and 2$\pi$ pulses is 95\%, respectively 80\%.
\label{fig1}}
\end{figure}
\end{center}

Rabi oscillations on the $\Delta n = 0$ carrier transition at a
frequency $\Omega_{{\rm carrier}} = 2 \pi\; 1090$\,kHz are shown
in Fig. 1a. For this, 100mW of light are focussed into a waist of
30$\mu$m at the position of the ion. We observe a contrast of more
than 95\%, the contrast being mainly limited by the thermally
distributed axial vibration mode ($\omega_x=2 \pi \,4.0$MHz),
which causes a spread in Rabi frequency owing to the
$n$-dependence as indicated in equation (\ref{Oxy}). We find that
our results are consistent with an average over a thermal phonon
distribution in the x-oscillator having $\langle n \rangle =
12(2)$.

\begin{figure}
\begin{minipage}{0.45\linewidth}
\begin{center}
\epsfxsize=0.9\linewidth \epsfbox{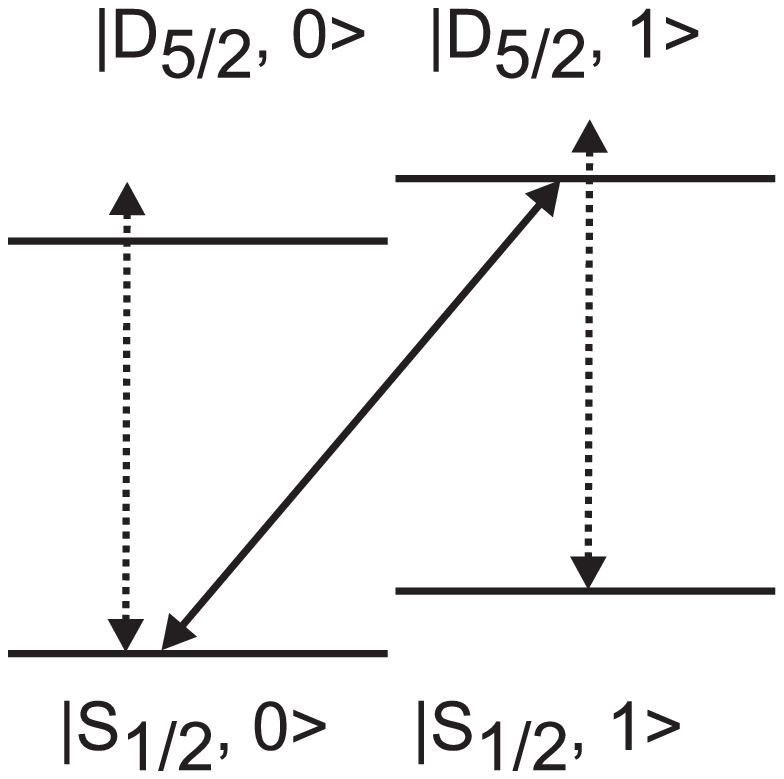}
\end{center}
\end{minipage}
\begin{minipage}{0.45\linewidth} \caption{Relevant energy levels: $|S_{1/2}, n=0\rangle$,
$|D_{5/2}, n=0\rangle$, $|S_{1/2}, n=1\rangle$, and $|D_{5/2},
n=1\rangle$. The solid arrow indicates the blue sideband
excitation, the dotted one the off-resonant carrier excitation.
\label{fig2}}
\end{minipage}
\end{figure}

\subsection{Sideband excitation}

The second building block is the sideband excitation, involving a
$\Delta n \ne 0$ transition, here performed on the $z$ vibration.
For convenience, we adopted the first blue rather than red
sideband. The equivalence of blue and red sidebands for these
studies was mentioned in section \ref{s_prelim}. The laser
frequency was red-detuned from the blue sideband resonance
(measured at low laser power) to compensate for the a.c. Stark
effect. The accuracy of the $\Delta n = +1$ operation is therefore
limited by off-resonant excitations on the $\Delta n = 0$
transition, as discussed in section \ref{redcorrect}, see also
Fig. 2. We indeed observe off-resonant carrier excitation, visible
as fast oscillations with a frequency near $\omega_z$ on top of
the $\Omega_{sideband}= 2\pi\,48$ kHz Rabi oscillation, see Fig.
3. Note that this frequency is far beyond the recoil frequency,
and thus the contrast shrinks to 75\% for a $\pi$ pulse.

\end{multicols}

\begin{center}
\begin{figure}[tbp]
\epsfxsize=0.95\hsize \epsfbox{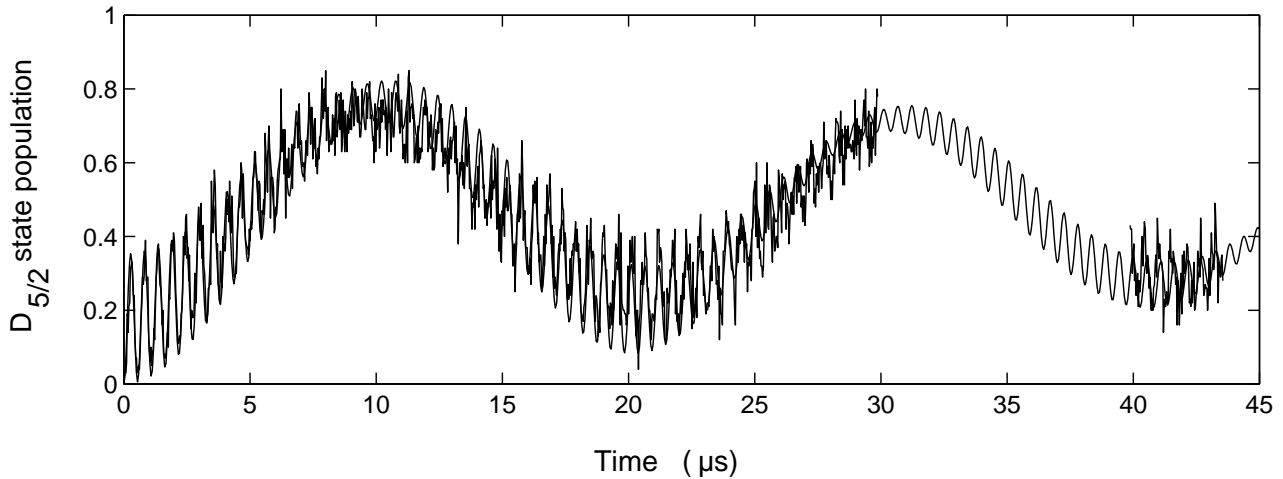} \caption{Rabi
oscillation on the blue sideband at 100mW excitation laser power.
See text for details. \label{fig3}}
\end{figure}
\end{center}

\begin{multicols}{2}

If the blue sideband $\omega_z$ is excited at a Rabi frequency of
$\Omega_{\rm sideband}= 2\pi\,7.4\,$kHz (Fig.\,1b), the
off-resonant carrier excitation is no longer detectable. A
contrast of 95\% for a $\pi$ pulse and 85\% for a $2\pi$ pulse is
measured, while equation (\ref{epsS}) predicts $\epsilon = 0.063$
and hence a fidelity $99.6$\% for the $\pi$ pulse. Under low-power
excitation, we do not observe any light shift.

%numerical simulations
The experimental data can be fitted by numerically solving the
Schr\"odinger equation in the truncated basis of the 4-level
system consisting of $|S_{1/2}, n=0\rangle$, $| D_{5/2},
n=0\rangle$, $|S_{1/2}, n=1\rangle$, and $|D_{5/2}, n=1\rangle$
states (see Fig. 2), and averaging over a thermal distribution of
population in the other vibrational modes. Three parameters in the
calculation were independently measured: the Rabi frequency on the
carrier $\Omega=2\pi 1090\,$kHz, the Lamb Dicke parameter
$\eta_z$, and the trap frequency $\omega_z = 2\pi\,1850$ kHz. The
detuning of the laser field $\delta$ is varied for optimum
contrast and we find excellent agreement between the data and a
numerical simulation for a value of $\delta_{{\rm th}}/2 \pi$=
-375\,kHz (Fig. 3). The discrepancy of 125\,kHz between
$\delta_{{\rm th}}$ and the experimentally determined value
$\delta_{{\rm exp}}/2\pi= -250$ kHz is probably caused by light
shifts due to the $D_{5/2}-P_{3/2}, S_{1/2}-P_{1/2}$, and
$S_{1/2}-P_{3/2}$ transitions. The sum of the calculated light
shifts from the dipole transitions is approximately 70\,kHz.

A third contribution to the overall light shift is expected from
the other radial ($y$) mode at 1.925\,MHz ($\delta
\omega/2\pi$=75\,kHz). Each level of the y-vibrational ladder is
light-shifted, resulting in a change in the $(n_y,n_y)(1_z,0_z)$
z-sideband resonance frequency by approximately $(\eta_y \cdot
\Omega_{carrier})^2 (n_y + 1)/(2(\omega_y-\omega_z)) =(n_y + 1)
(2\pi)\;0.8$\,kHz. Assuming a mean vibrational quantum number of
25, a 20\,kHz spread in shifts is expected. To model this
situation we solved the Schr\"odinger equation for different
$y$-oscillator occupations and averaged the results over a thermal
distribution (see Fig. 4). The loss of contrast at $\Omega t =\pi$
is dominated by this thermal averaging, rather than laser
intensity and magnetic field fluctuations.

To compare the experimental findings with an optimum $\pi$ pulse,
for a given pulse length, we investigated numerically the
parameter space and found that an optimum $\pi$ pulse (92\%
contrast of Rabi $|S_{-1/2} \rangle \leftrightarrow |D_{-5/2}
\rangle$ oscillation) could be achieved for a detuning of
$\delta_{{\rm th}}/2\pi= -355$\,kHz. We conclude that the detuning
chosen in the experiment was off by 20\,kHz. Note, that the
electron shelving detection method only measures the internal
state occupation, {\em not} the vibrational one. Hence, the
measured contrast does not directly yield the fidelity. Equation
(\ref{epsS}) predicts $\epsilon = 0.41$ for this case, and hence a
fidelity of 83\%. The 20\,kHz deviation from the ideal detuning
leads to a 5\% loss in contrast and the thermally distributed
$y$-vibration mode with $\langle n\rangle = \langle 25 \rangle$
further reduces the contrast to 75\%. The corresponding fidelity
was near 64\%. For the details of the $\pi$ pulse dynamics, see
Fig.4.

\begin{center}
\begin{figure}[tbp]
\epsfxsize=0.95 \hsize \epsfbox{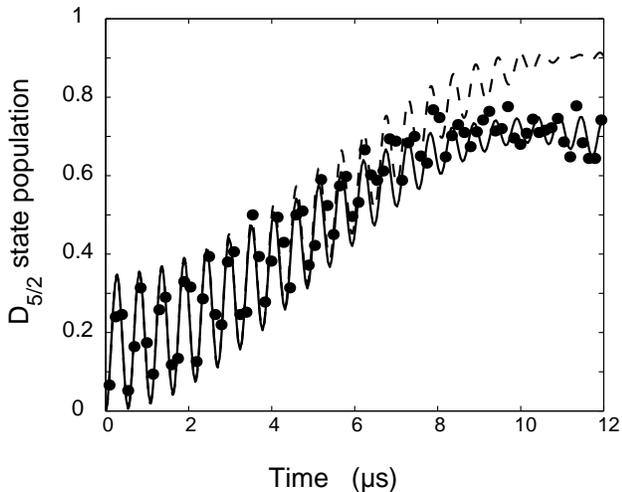} \caption{$\pi$
pulse on the blue sideband, detailed view. The data of Fig. 3 are
plotted together with a fit taking into account a thermally
distributed radial $y$-mode (see text for the set of parameters).
The dotted line gives a prediction for optimized parameters, where
all relevant vibration modes are assumed to be cooled to the
ground state, and the detuning is set for maximum contrast. A
contrast of 92\% is predicted under these optimum conditions, and
the fidelity $1-\epsilon^2$ given by equation (\protect\ref{epsS})
would then be 83\%.
\label{fig4}}
\end{figure}
\end{center}

To conclude, we have presented a theoretical discussion of two
types of quantum gate which couple the internal ion state with
its motion, and an experimental study of one of these. Our
calculations lead to quantitative statements of the precision of
the operations, taking into account the complete Hamiltonian
including all the vibrational states and the off-resonant coupling
terms. These statements are given in equations (\ref{epsC}),
(\ref{TC2min}) for the Monroe {\em et al} ``magic Lamb Dicke
parameter" gate, and in equations (\ref{epsS}), (\ref{TSmin}) for
the Cirac-Zoller {\sc swap} gate. If we set a limit of a fixed
number of laser
wavelengths for the spacing of the ions in a linear trap, then we
arrive at a gate time per ion, for the {\sc swap} gate, which
depends only on the precision and the choice of ion and
transition. This time is given in equation (\ref{TSmm}) and table
\ref{t_rate}.

In our experimental studies we have demonstrated a contrast of
75\% for a $1/T_S = 100\,$kHz {\sc swap} operation, and 95\% for a
$1/T_S = 14\,$kHz {\sc swap} operation on a single trapped ion.
These gate rates are respectively 21 and 3 times the relevant
recoil frequency. In both cases the contrast could be
significantly improved by cooling a further vibrational mode.

\bibliographystyle{prsty}
\bibliography{myrefs,quinforefs}

\begin{thebibliography}{10}

\bibitem{95:Cirac}
J.~I. Cirac and P. Zoller, Phys. Rev. Lett. {\bf 74},  4091  (1995).

\bibitem{95:Monroe}
C. Monroe {\it et~al.}, Phys. Rev. Lett. {\bf 75},  4714  (1995).

\bibitem{98:Turchette}
Q.~A. Turchette {\it et~al.}, Phys. Rev. Lett. {\bf 81},  3631  (1998).

\bibitem{97:SteaneB}
A.~M. Steane, Appl. Phys. B {\bf 64},  632  (1997).

\bibitem{97:Monroe}
C. Monroe {\it et~al.}, Phys. Rev. A {\bf {55}},  R2489  (1997).

\bibitem{99:Roos}
C. Roos {\it et~al.}, Phys. Rev. Lett. {\bf 83},  4713  (1999).

\bibitem{98:James}
D.~V.~F. James, Applied Physics B {\bf 66},  181  (1998).

\bibitem{85:Deutsch}
D. Deutsch, Proc. R. Soc. Lond. A {\bf 400},  97  (1985).

\bibitem{96:Barenco}
A. Barenco, Contemporary Physics {\bf 37},  375  (1996).

\bibitem{98:Steane}
A. Steane, Rep. Prog. Phys. {\bf 61},  117  (1998).

\bibitem{99:Sorensen}
A. Sorensen and K. Molmer, Phys. Rev. Lett. {\bf 82},  1971  (1999).

\bibitem{99:Molmer}
K. Molmer and A. Sorensen, Phys. Rev. Lett. {\bf 82},  1835  (1999).

\bibitem{00:Sorensen}
A. Sorensen and K. Molmer,   (2000), quant-ph/0002024.

\bibitem{98:WinelandB}
D.~J. Wineland {\it et~al.}, J. Res. Natl. Inst. Stand. Technol. {\bf 103},
  259  (1998).

\bibitem{96:Hughes}
R.~J. Hughes {\it et~al.}, Phys. Rev. Lett. {\bf 77},  3240  (1996).

\bibitem {comment}
During preparation of this paper, we learned of the work D. Jonathan, M. B.
Plenio and P. L. Knight,
``Fast quantum gates for cold trapped ions", quant-ph/0002092, which
proposes a method to gain access to the higher speed with travelling wave excitation by making use of
the light shift.

\bibitem{99:Nagerl}
H.~C. N\"{a}gerl {\it et~al.}, Phys. Rev. A {\bf 60},  145  (1999).

\bibitem{86:Sauter}
T. Sauter, W. Neuhauser, R. Blatt, and P.~E. Toschek, Phys. Rev. Lett. {\bf
  57},  1696  (1986).

\bibitem{86:Nagourney}
W. Nagourney, J. Sandberg, and H. Dehmelt, Phys. Rev. Lett. {\bf 55},  2797
  (1986).

\bibitem{86:Bergquist}
J.~C. Bergquist, R.~G. Hulet, W.~M. Itano, and D.~J. Wineland, Phys. Rev. Lett.
  {\bf 57},  1699  (1986).

\bibitem{Roos2000}
C. Roos, Ph.D. thesis, 2000, unpublished.

\end{thebibliography}

\end{multicols}
\end{document}